\documentclass[a4paper,10pt,aps]{revtex4-1}

\usepackage{amsmath}
\usepackage{bbm}
\usepackage{graphicx,amssymb,bm,latexsym,color,epsf}
\usepackage{breqn}
\usepackage[colorlinks]{hyperref} %nice refs and links

\newcommand{\be}{\begin{equation}}
\newcommand{\ee}{\end{equation}}
\newcommand{\bea}{\begin{eqnarray}}
\newcommand{\eea}{\end{eqnarray}}

\makeatletter
\let\cat@comma@active\@empty
\makeatother

\begin{document}
%%%%%%%%%%%%%%%%%%%%%%%%%%%%%%%%%%%%%%%
%%%%%%%%%%%%%%%%%%%%%%%%%%%%%%%%%%%%%%%%%%%%%%%%%%%%%%%%%
\title{Ultraviolet properties of 
Lifshitz-type scalar field theories}
%%%%%%%%%%%%%%%%%%%%%%%%%%%%%%%%%%%%%%%%%%%%%%%%%%%%%%%%%

\author{Dario Zappal\`a}
\email{dario.zappala@ct.infn.it}
\affiliation{INFN, Sezione di Catania, via S. Sofia 64, I-95123, 
Catania, Italy}

\begin{abstract}
\vskip 30pt
\centerline{ABSTRACT}
\vskip 10pt
We consider Lifshitz-type scalar field theories that exhibit anisotropic scaling laws 
near the ultraviolet fixed point, with explicit breaking of  Lorentz symmetry. 
It is shown that, when all momentum dependent vertex operators are discarded, 
actions with  anisotropy parameter $z=3$ in 3+1 dimensions generate  Lorentz symmetry 
violating  quantum corrections that are suppressed by inverse powers of the momentum, 
so that the symmetry is sensibly restored in the infrared region. In the ultraviolet region,
the singular behavior of  the  corrections is strongly  smoothened: only logarithmic 
divergences show up, producing  very small changes of the couplings 
over a range of momentum of many orders of magnitude.  In the particular case where 
all couplings are equal, the theory  shows a Liouville-like potential and quantum corrections 
are exactly summable, giving an asymptotically free theory. 
However, the observed  weakening of the  divergences is not sufficient to 
avoid a residual fine tuning of the mass parameter at a very high energy scale, in order to 
recover a physically acceptable mass in the infrared region.

\end{abstract} 
\maketitle

\setcounter{page}{2}

%%%%%%%%%%%%%%%%%%%%%%%%%%%%%%%%%%%%%%%%%%%%%%%%%%%%%%%%%%%%%%%%%%%%
\section{Introduction}
%%%%%%%%%%%%%%%%%%%%%%%%%%%%%%%%%%%%%%%%%%%%%%%%%%%%%%%%%%%%%%%%%%%%%%

Understanding the  ultraviolet (UV) structure of field theories is an essential issue to probe the mathematical consistency of the theory itself, but also to explore 
possible extensions or completions of models, such as the Standard Model of the elementary particle interactions and its interplay with General Relativity.
Along this line, much effort has been devoted to find a way to smoothen the severe divergences that show up for instance in scalar theories, which are not 
protected by gauge symmetry. In particular, the inclusion of higher derivative operators has the effect of reducing the degree of 
UV divergence of the diagrams. This  approach has the desired property of improving the renormalizability of the theory \cite{thirring,Pais:1950za},
and has been studied at length, both in the context of quantum field theories and in gravity 
\cite{Stelle:1976gc,deUrries:1998obu,deUrries:2001txi,Hawking:2001yt,Rivelles:2003jd,Smilga:2004cy,Kruglov:2006gu,Anselmi:2007ri}.

In this regard, a helpful insight comes from condensed matter physics, where general properties of higher derivative theories have been 
related to the  presence of Lifshitz points, i.e. points in the phase diagram at which it is observed the co-existence of 
a disordered phase, a spatially  homogeneous ordered phase and a spatially modulated  ordered phase,
and their appearance is typically related to a particular balance between the standard two derivative term and a higher derivative term in the action
\cite{Horn,hornrev,selke1988,Diehl,Diehl3}. 
Then, phase transitions associated to Lifshitz points can be put in relation with  continuum limit and renormalizability properties  of the related 
higher derivative  euclidean field theory.

In general, condensed matter systems admit the existence of isotropic and anisotropic Lifshitz points as well, i.e. points that maintain full rotational symmetry on all 
coordinates, the former, and points that present anisotropic scaling  properties and are  described by two different correlation lengths in different space directions,
the latter.  Accordingly, isotropic points require a symmetric structure of the higher derivative terms that involves the same order of derivatives for each coordinate, 
while  anisotropic points are characterized by different number of derivatives on different coordinates.  Therefore, properties of the transitions and dependence on 
dimensionality are different in the two cases. 

Although the isotropic case possesses  a rich phase structure \cite{Horn,Diehl3,Bonanno:2014yia,Zappala:2017vjf,Zappala:2018khg,Zapp,Defenu}, 
only the anisotropic case  can meaningfully  be connected with field theories defined in Minkowski space, that are of interest in the high energy context.
In fact, the presence of time derivatives of order larger than two directly leads to the Ostrogradski instability, corresponding to  Hamiltonians unbounded 
from below with  violation of unitarity \cite{deUrries:1998obu,Woodard:2006nt}. To avoid these complications, it is unavoidable to resort to anisotropic Lifshitz systems 
where the time derivative is of second order, while in the spatial sector higher derivative terms  are present 
(usually the anisotropy parameter $z$ is introduced to  specify  the order of  the spatial derivatives as equal to $2z$).

Even in anisotropic form, the Lifshitz points have the effect of enhancing the UV sector of the theory, reducing the degree of divergence.
The price to pay is  the breaking of Lorentz invariance  at high energy scales, and it becomes essential to verify that 
Lorentz symmetry  becomes manifest in the infrared (IR) region \cite{Chadha:1982qq}, due to a suppression of the Lorentz breaking terms
 below the level of the experimental limits.

This kind of anisotropic models received much attention, especially after the introduction of the Ho\v{r}ava-Lifshitz formulation of the gravitational 
theory \cite{horava} with the aim of deriving a renormalizable approach to quantum gravity, 
and a lot of work has followed in various contexts ranging from gravity, black holes, cosmology 
\cite{Cai,Brand,Takahashi:2009wc,Kiritsis:2009sh,Calcagni:2009ar,Son,Eune:2010,cognola2016,Barv1,Barv2,Barv3,Barv4},
to gauge, scalar or fermionic  field  theory \cite{horava:ym,
Iengo,Dhar:2009dx,Kikuchi,eune,alexandre,Chao:2009tqf,Solomon:2017nlh}.
In particular, in \cite{Iengo} and subsequently in \cite{alexandre} it was pointed out that anisotropic scalar theories which include all possible renormalizable
operators (according to the Lifshitz scaling), have the serious drawback of producing large Lorentz violating quantum corrections that are 
not compatible with the experimental observations, unless a quite severe fine tuning of various couplings in the UV region is performed.

In this paper, we reconsider this problem and select a specific class of scalar theories that avoids the occurrence of the large Lorentz violating corrections.
We shall focus on a $3+1$-dimensional theory with  anisotropy parameter $z=3$,  because of its particular Lifshitz point structure and  then, 
after reducing the bare action to momentum independent vertices only, we  analyze the generated quantum corrections.  Since Lorentz symmetry is 
absent from the beginning, we are allowed to treat divergent terms by means of a non-Lorentz invariant regulator, namely a  three-momentum cut-off and,
due to the very simple nature of the divergent diagrams, we are able to carry out our computations
without resorting to the rotation to Euclidean coordinates, in the spirit of  \cite{Farias:2011aa,Bonanno:2020thj}. Within this approach the main interesting properties of the 
renormalized theory are pointed out and the crossover from the UV to the IR region is analyzed.

In Section \ref{sec2}, we  study the properties of  Lifshitz scaling of higher derivative scalar field theories and their consequently modified renormalizability.
In Section \ref{sec3}, we analyze one specific model with $z=3$ and its diagrammatic expansion, 
while in  Section \ref{sec4}  the  RG flow  of the various couplings  in the UV and  IR regimes. The conclusions are reported in Section \ref{sec5}. 

%%%%%%%%%%%%%%%%%%%%%%%%%%%%%%%%%%%%%%%%%%%%%%%%%%%%%%%%%%%%%%%%%%%%
\section{Lifshitz scaling}
\label{sec2}
%%%%%%%%%%%%%%%%%%%%%%%%%%%%%%%%%%%%%%%%%%%%%%%%%%%%%%%%%%%%%%%%%%%%%%

We are specifically interested in the following Minkowskian Action
\begin{equation}
S=\!\int\!d^{3}x\,dt\left(
\frac{1}{2}
\left(\partial_{t}\phi\right)^{2}-
\frac{ \widehat a_{z}}{2}  \left(\partial_{i}^{z}\phi\right)^{2}-
\frac{ \widehat a_{z-1}}{2}  \left(\partial_{i}^{z-1}\phi\right)^{2}-...-
\frac{ \widehat a_{1}}{2}  \left(\partial_{i}\phi\right)^{2}-V\right),
\label{eq: actionorig}
\end{equation}
%%%%
where we included higher derivatives  of the spatial coordinates labelled by the index 
$i$ (and $i$ is summed over $i=1,2,3$), up to order $2z$ (with $z>1$).
Hatted couplings indicate that they are dimensionful quantities.
This is a particular case of the general problem of anisotropic Lifshitz scaling where the full set of Euclidean coordinates is 
composed of two subsets which have different scaling properties in proximity of a Lifshitz point. In the case of Eq. (\ref{eq: actionorig}),
one subset is represented by the time coordinate and the other by the 3 spatial coordinates, with scaling dimensions respectively 
$[t]_{s}=-z$ and $[x^{i}]_{s}=-1$, so that for a scale transformation with rescaling parameter $b>1$, one observes the following  non-uniform 
scaling of space and time
%%%%%%%%%%%%
\begin{equation}
\label{eq:scaling}
t\to b^{z}\,t, \;\;\;\;\;\;\;\; x^{i}\to b\,x^{i}.
\end{equation}
%%%%%%%%%%%%%%%%
This leads to a substantial difference with respect to the standard scaling
associated  with the canonical dimensions of the  parameters entering Eq. (\ref{eq: actionorig}) 
which occurs in proximity of the gaussian fixed point. 

So, for instance, the coefficients  $ \widehat a_{j}$ (with $j>1$)
of the higher derivative terms  appearing in Eq. (\ref{eq: actionorig}),  
have negative canonical dimension and, accordingly, the corresponding terms are  perturbatively non-renormalizable. 
However the presence of another fixed point, the Lifshitz point, ensures different scaling rules that redefine the 
renormalization  properties of the various terms and, in accordance with the scaling rules in Eq. (\ref{eq:scaling}), we
derive the scaling dimension of the field from the scaling properties of the term proportional to $\widehat a_{z}$ 
(which is assumed to have scaling dimension $[\widehat a_z]_{s}=0$) : 
%%%%%%%%%%%%
\begin{equation}
\label{eq:fielddim}
[\phi]_{s}=\frac{3-z}{2} \;.
\end{equation}
%%%%%%%%%%

Actually, this is a particular case of a more general analysis of a $d$-dimensional system with  $d_s$ dimensions in one subset of 
coordinates,  instead of the 3 spatial dimensions,  and $d-d_s$ dimensions in the other subset, instead of the only time dimension. 
In fact, in general by allowing for an anomalous dimension $\eta$
of the $d_s$-dimensional subset, i.e. by redefining  the scaling dimension  $[\widehat a_z]_{s}=-\eta$, one obtains $[\phi]_{s}=[z\,(d-d_s-2)+d_s+\eta]/2$, that clearly 
reproduces Eq. (\ref{eq:fielddim}) for  $d=3+1$, $d_s=3$ and $\eta=0$. Nevertheless, for our purposes of analyzing a (3+1)-dimensional Minkowskian 
space-time theory, it  is sufficient to retain the scaling rule in Eq. (\ref{eq:fielddim}) with zero anomalous dimension.

 Once the scaling dimension of the field is known, we can investigate the constraints on the various parameters that characterize the  existence of a non-trivial Lifshitz
 point and,  in general, this means finding the upper and  lower critical dimension for the action considered. In our case, where the dimensions are kept fixed, this 
 reduces to a constraint on $z$. In fact, by considering the potential in Eq. (\ref{eq: actionorig}) as an expansion in powers of the field
%%%%%%%%%
\begin{equation}
{ V}=\sum^{\overline n}_{n=2}
\frac{\widehat g_n}{n!}\phi^{n}
\;\; ,
\label{eq: poten}
\end{equation}
%%%%%%%%%%%%%%
we easily derive the scaling dimension of the generic $\widehat g_n$:
%%%%%%%%%%%%
\begin{equation}
\label{eq:gdim}
[\widehat g_n]_{s}=\left(1-\frac{n}{2}\right) \,(z +3 ) +nz 
\end{equation}
%%%%%%%%%%
and therefore, the marginality condition $[\widehat g_n]_{s}=0$ corresponds to the constraint 
%%%%%%%%%%
\begin{equation}
\label{eq:marginal}
\left(2+n \right) \,z = 3 \left(n -2 \right)\; .
\end{equation}
%%%%%%%%%%%%
From Eq. (\ref{eq:marginal}), we find that, when $z=1$, the marginal coupling corresponds to $n=4$, as is well known,
and when $z=2$ to $n=10$, and when z=3 no finite $n$ satisfies Eq. (\ref{eq:marginal}) as $[\widehat g_n]_{s}=6$ for any $n$.

By inverting the argument, if we require the marginal coupling to be $\widehat g_4$ so that, at least for theories with $Z_2$ symmetry 
(or $O(N)$ symmetry for a multicomponent field), no term  in the potential (\ref{eq: poten})  
has positive scaling dimension (apart from the quadratic mass term), we find that in Eq. (\ref{eq:marginal}), $[\widehat g_4]_{s}=0$ implies $z=1$. This indicates that 
our 3+1 dimensional system  with $z=1$ corresponds to the upper critical dimension which instead becomes larger at  larger values of $z$. 
On the other hand, the lowest dimension below which the non-trivial Lifshitz point disappears, corresponds to the vanishing of the 
scaling dimension of the field.  According to  Eq. (\ref{eq:fielddim}), for our 3+1 dimensional system this  implies $z=3$.

The above analysis indicates that for a $O(N)$ theory in 3+1 dimensions, a  non-trivial Lifshitz point is expected  for $z=2$ only, while when $z=1$ and $z=3$ we are 
respectively at the  upper and lower critical dimension of the system where  only the trivial Lifshitz point solution of the Renormalization Group analysis is found.
Namely, it corresponds  to the quadratic  action with, respectively, $z=1$ and $z=3$:
%%%
\begin{equation} 
S_{FP}=\!\int\!d^{3}x\,dt\left(\frac{1}{2} \left(\partial_{t}\phi\right)^{2}-
\frac{\widehat a_{z}}{2}  \left(\partial_{i}^{z}\phi\right)^{2}
\right) \;.
\label{eq: lpaction}
\end{equation}
%%%%

These results are totally in agreement with those of \cite{hornrev,Diehl,Paris:2017buq} where the existence of a Lifshitz point solution is 
investigated for  generic number of dimensions of the two subsets of coordinates and with $z=2$ fixed. 
Clearly, if $z=1$ in Eq. (\ref{eq: lpaction}), we are left with the standard Gaussian fixed point, while $z=3$ yields a 
non-standard higher derivative action that we expect to be responsible for a well behaved UV sector of the theory.

In fact, it is easy to realize that when $z=3$, the scaling dimension of the field vanishes according to Eq. (\ref{eq:fielddim}),
and therefore all  possible interaction terms (and mass term) that can be added to $S_{FP}$ have couplings with positive 
scaling dimension, which means that these operators are relevant with respect to to this fixed point solution and get suppressed in the UV regime.
This is in agreement with the naive expectation that the UV divergences of such  a higher derivative theory with $z=3$ are strongly (if not totally)
suppressed by the enhanced power of the momentum  ($k^6$ instead of $k^2$) in the propagator. 

Therefore, since we want to investigate how much the fixed point solution (\ref{eq: lpaction}) does protect from the UV divergences,
we shall consider the most favourable case with $z=3$. Moreover, for  $z=2$,
the appearance of the non-trivial Lifshitz point in addition to the one of (\ref{eq: lpaction}), does  certainly modify the structure  of Renormalization Group flow
and this could generate some complication in the renormalization  process. 
On the other hand, larger values of $z$, such as $z=4$, imply a negative scaling dimension of the field, thus producing an unstable theory. 

In conclusion we will focus on the action in Eq. (\ref{eq: actionorig}) with $z=3$ and the potential $V$ specified in Eq. (\ref{eq: poten}), with generic $\overline n$. 
In fact, as it will be shown next, the divergence of  the loop diagrams of this theory does not depend on the index $n$ that defines the number of legs of each vertex. 

We turn now to the analysis of the degree of divergence of the various loop diagrams of such a theory and, since we are including in the action (\ref{eq: actionorig}) 
an explicit violation of the Lorentz symmetry, we are allowed  to treat  space and time coordinates separately in the loop integrals. Then, we find convenient to 
maintain Minkowskian coordinates, thus avoiding the rotation to Euclidean coordinates, and to first perform the integration on the energy component in the 
four-momentum integral along the Feynman contour, so that the divergences appear only  in the integration on the three momentum. We shall take care 
of them by means of a Lorentz violating three-momentum cut-off $\Lambda$.

In order to compute the degree of divergence of a generic diagram, we recall the well known fact that each propagator can be split into the product of two factors,
with one of the two being a pole of the energy integral. 
So for instance, in a one loop integral with two propagators we end up with the product of four terms, two of which give poles  that are relevant in the 
Feynman contour integration :
%%%%%%%%%%
\begin{eqnarray}
&&R_1(p)=\int\!\frac{d^{3}k  \; d k_0 }{(2\pi)^{4}}
\; \frac{1}{
\left[  k_{0}^{2} -A^{2}+ i \epsilon    \right ] }
\;
\frac{1}{ \left[     (p_0+k_{0})^{2} -B^{2}+ i \epsilon \right ] }= \nonumber \\
&&\int\!\frac{d^{3}k  \; d k_0 }{(2\pi)^{4}}
\; \frac{1}{ \left[  k_{0} -A+ i \epsilon    \right ] \left[  k_{0} + A - i \epsilon    \right ] }
\;
\frac{1}{\left[ (p_0+k_{0}) -B+ i \epsilon \right ]  \left[ (p_0+k_{0})+B- i \epsilon \right ] } \; ,
\label{eq: poles}
\end{eqnarray}
%%%%%%%%%%%%%
where we defined, according to the action (\ref{eq: actionorig}) with $z=3$ and the potential (\ref{eq: poten}), 
%%%%%%%%%%
\begin{equation}
A=\sqrt{\widehat a_3 \, \vec{k}^{\,6}     + \widehat a_2 \, \vec{k}^{\,4} + \widehat a_1 \, \vec{k}^{\,2} + \widehat g_2}\;\;\; ,  \;\;\;
B=\sqrt{\widehat a_3 \, ( \vec{p} +\vec{k})^{\,6}      +\widehat a_2 \,  (\vec{p} +\vec{k})^{\,4} +\widehat a_1 (\vec{p} +\vec{k})^{\,2} +\widehat g_2} \;\; .
\label{eq:definitions}
\end{equation}
%%%%%%%%%%%%%%%%%%%%
We then perform the integration over $k_0$ by reducing the double pole to a sum of simple poles, according to the expression:
%%%%%%%%%
\begin{equation}
\frac{1}{x-P_1} \;\; \frac{1}{x-P_2} = \left ( \frac{1}{x-P_1} - \frac{1}{x-P_2} \right )  \frac{1}{(P_1 - P_2) } \;,
\label{eq: linear}
\end{equation}
%%%%%%%
where $x$ indicates the integration variable and $P_1,\,P_2$ are $x$-independent expressions.

The outlined procedure can be extended  also to multiloop diagrams and this allows us to infer the rule to count the degree of divergence of each diagram. 
In fact, after performing the integrations on the energy component of every loop, we find for
generic $z$ (which is then set to $z=3$  in our model) that each loop brings $3$  powers of momentum  
from the three-momentum integration,  while each internal line contributes with $- 2z$ powers of momentum, with the exception of one 
internal line for each loop that contributes with a power equal to $-z$, because of the cancellation produced by  the integration over $k_0$.  
With the help of this rule we compute the degree of divergence of 
a diagram (i.e. the powers of the three-momentum cut-off $\Lambda$ ) $D_\Lambda$: 
%%%%%%%%%%%
\begin{equation}
D_\Lambda =  \left [ 3L - 2z\,(I-L)  - z\,L \right ]|_{z=3} \;= \, 6 (L - I)
\label{eq: diver}
\end{equation}
%%%%%%%%%%%%%%%%%
where $L$ is the number of loops and $I$ the number of internal lines of the diagram 
(below, we shall  label the number of external lines as $E$ 
 and  the number of vertices with $n$ lines as $V_n$).

If we now recall the well known relations due to the diagram topology,
 $L=I- \sum_n V_n +1$ and  $\sum_n (n\,V_n )=E +2I$, 
 where the sum is understood over the kind of vertices entering the diagram, we obtain from Eq. (\ref{eq: diver}) :
%%%%%%%%%%%
\begin{equation}
D_\Lambda = \left [ \left ( \sum_n (n V_n)  - E \right ) \frac{(3-z)}{2} + (3+z) \left ( 1-\sum_n V_n \right ) \right]  \Bigg |_{z=3}  \;=  \, 6\, \left (1 - \sum_n V_n   \right) \; .
\label{eq: diver2}
\end{equation}
%%%%%%%%%%%%%%%%% 
Incidentally, the same conclusions would have been obtained had we inspected the divergence structure of the theory after Wick rotation to Euclidean coordinates 
and adopted a symmetric cut-off on the modulus of the four-momentum in the loop integrals.
 
It is now evident that, in the presence of modified  'higher derivative' propagators with $z=3$, only logarithmic divergences show up in diagrams containing only one
vertex, whatever is the value of $n$ and the number of external legs $E$. Consequently the UV sector of such a theory turns out to be rather simple
and we shall analyze it in  Section \ref{sec3}.

One could object that the action (\ref{eq: actionorig}) with $z=3$ does not include all renormalizable terms,
because the scaling dimension of the field is vanishing, as shown in Eq. (\ref{eq:fielddim}).  
Therefore, besides including any possible  value of $n$,  which characterizes the vertex in the potential (\ref{eq: poten}), it is admissible to include also vertices like 
%%%%%%%%%
\begin{equation}
w_{m,s} \;
\phi^m  \left(\partial_{i}^{s}\phi \partial_{i}^{s}\phi   \right) \; , 
\label{eq:undesired}
\end{equation}
%%%%%%
with  positive integer  $m$  and  $s=1, 2, 3$, still maintaining non-negative 
scaling dimension of the corresponding coupling (higher values of $s$  would instead yield  negative scaling dimension). Clearly, such kind of vertex
that brings a dependence on the  momentum of two lines, does increase the degree of the divergence when the momentum of these two lines 
involves the integrated internal  momentum of a loop, and therefore it requires a more accurate rearrangement of all the diverging integrals in 
the renormalization procedure; but nevertheless the renormalizability property guarantees the full cancellation of divergences by suitable 
insertion of counterterms.

 On the other hand, there is one important property  concerning the class of vertices in Eq. (\ref{eq:undesired})  that must be considered.
Namely, the requirement  $w_{m,s} =0$,  for all $m>0$ and $s=1, 2, 3$ in the bare action (\ref{eq: actionorig})  of our model,  implies the 
absence of any UV divergence in the following expansion of the $(m+2)$-point Green functions generated by the action (\ref{eq: actionorig}), 
with $m$ out of the $(m+2)$  momenta set to zero: 
\begin{equation}
(p_i \, p'_i)^s \;\,{\cal W}_{m,s} = (p_i \, p'_i)^s \, \Bigg[ \Bigg(  \frac{\partial^2}{\partial p_j  \; \partial p'_j  } \Bigg )^s \Gamma^{^{(m+2)}}(p,p',0,...,0) \Bigg ]_{p=p'=0} \; .
\label{momgreen}
\end{equation}

The relevance of this property is clear. In fact  any UV divergence 
in  terms like  ${\cal W}_{m,s}$  in Eq. (\ref{momgreen}) 
could not be cancelled, because this is achievable only  by means of a counterterm generated by 
the corresponding vertex in (\ref{eq:undesired}) that, however, 
is missing  in the bare action (\ref{eq: actionorig}) where we set all $w_{m,s} =0$.

To prove the above statement, we notice that, without  vertices  (\ref{eq:undesired}) in the bare action,
the only way to get an  explicit dependence on the external momenta to the power $2s$,
analogous to  the $2s$ derivatives in Eq. (\ref{eq:undesired}), is given by  the expansion in Eq. (\ref{momgreen}). 
In addition,  it is evident that the  external momenta dependence of a Green function of the action (\ref{eq: actionorig}), 
which is essential  to obtain  non-vanishing derivatives in Eq. (\ref{momgreen}),
cannot be generated by diagrams with one vertex only, but comes instead from diagrams containing at least two vertices.

Then, with  all $w_{m,s} =0$ in (\ref{eq: actionorig}),  $D_\Lambda$ must be read  from Eqs. (\ref{eq: diver}) and (\ref{eq: diver2}),
and, as  we had already observed, it implies that only diagrams with one vertex are UV divergent. 
Therefore, Green functions that carry external momentum 
dependence and consequently contain at least two vertices, are finite. 
This is already  sufficient to demonstrate our statement but, since  ${\cal W}_{m,s}$ is the result of $2s$ derivatives 
with respect to the  external momenta of such a  Green function, 
we have a further increase of the degree of UV convergence of ${\cal W}_{m,s}$, as compared with $\Gamma^{^{(m+2)}}(p,p',0,...,0)$.  

We conclude that the constraint  $w_{m,s} =0$ for all $m>0$ and $s=1, 2, 3$ 
is sufficient to warrant only UV finite contributions to ${\cal W}_{m,s}$, which are harmless in the analysis of the renormalized theory,
in much the same way as the $g_4\,\phi^4$ vertex in the action of a standard renormalizable theory in four dimensions generates
only UV finite $\Gamma^{(n)}$ Green functions with $n\geq 6$, 
while the inclusion of the $g_6\,\phi^6$ vertex, which produces  unmanageable UV divergences,
would  instead spoil the UV convergence of those Green functions.

At this point, we recall that  momentum dependent vertices in Eq. (\ref{eq:undesired})
are responsible of another  undesired feature, as  shown in \cite{Iengo}.
In the latter, a 4+1 dimensional model  with $z=2$ is studied,  but  the effect produced by these 
momentum dependent vertices is essentially the same in their case and in ours.
More specifically, if  we compute the one loop correction to the coefficient $a_1$,  
generated  by the term  $\phi^2  \left(\partial_{i}\phi  \partial_{i}\phi \right)$, of the kind displayed in Eq. (\ref{eq:undesired}), 
suitably included  into the action  (\ref{eq: actionorig}), we get logarithmic corrections to $a_1$.
Then, if we compare the logarithmic corrections obtained for two different kind of fields with different couplings, we find 
that  the renormalized  $a_1$ of the two fields differ by a logarithmically growing quantity when the IR region is approached,
unless a fine tuning of the couplings of the two fields in the UV region is enforced. This discrepancy, as discussed  in \cite{Iengo},
is certainly not compatible with the  observed Lorentz invariance in the IR region, that requires instead equal renormalized  $a_1$ 
for the two fields, at least within the experimental errors.
(For a detailed review about  IR effects of Lorentz invariance violation introduced 
at extremely large energy scales and, in particular, with reference  to quantum gravity models, see
\cite{Collins1,Collins2}).

As these logarithmic corrections, unacceptable on phenomenological grounds,  disappear if the momentum dependent vertices in 
Eq. (\ref{eq:undesired}) are omitted from the full action, leaving only finite power law corrections to the dispersion relation 
(as will be discussed in  Section \ref{sec3}), we choose to bypass the renormalizability criterion that would instead require 
the inclusion of the terms displayed in Eq. (\ref{eq:undesired}). 
In other words, the action in (\ref{eq: actionorig})  analyzed in Section \ref{sec3}, contains all renormalizable operators, according to the 
Lifshitz scaling around the fixed point solution (\ref{eq: lpaction}), with the exception of momentum dependent vertices; the couplings associated to 
the latter will be  set to zero in the bare action.
The evidence that, under these assumptions,
${\cal W}_{m,s}$  in Eq. (\ref{momgreen}) gets only UV finite corrections,  guarantees the consistency of our choice.  
%%%
 %%%%%%%%%%%%%%%%%%%%%%%%%%%%%%%%%%%%%%%%%%%%%%%%%%%%%%%%%%%%%%%%%%%%
\section{Renormalized theory}
\label{sec3}
%%%%%%%%%%%%%%%%%%%%%%%%%%%%%%%%%%%%%%%%%%%%%%%%%%%%%%%%%%%%%%%%%%%%%%

According to the issues considered in Section \ref{sec2}, we restrict our analysis to the action
%%%
\begin{equation}
S=\!\int\!d^{3}x\,dt\left[
\frac{1}{2}
\left( \partial_{t}\phi\right)^{2}-
\frac{ \widehat a_{3}}{2}  \left(\partial^2\partial_{i}\phi  \partial^2\partial_{i}\phi  \right)-
\frac{ \widehat  a_{2}}{2} \left(\partial^2\phi  \partial^2\phi  \right)-
\frac{ \widehat a_{1}}{2}  \left(\partial_{i}\phi  \partial_{i}\phi  \right)    -V(\phi) \right],
\label{eq: actioneff}
\end{equation}
%%%%
with $V$  given in Eq. (\ref{eq: poten}) and where, in order to keep the number of couplings finite, 
we choose the maximum power of the field, $\overline n$, finite.
 Note that any other term, quadratic in the field and containing 2 or 4 or 6 derivatives but with different 
derivative ordering with respect to Eq. (\ref{eq: actioneff}), like e.g. $\left(\partial_i\partial_{j}\partial_k \phi \; \partial_i\partial_{j}\partial_k \phi   \right)$
or $\left(\partial_i\partial_{j}\phi \; \partial_i\partial_{j} \phi   \right)$, is always reducible, after integration by parts, to
 one of the  terms in (\ref{eq: actioneff}).

 It is now convenient to redefine the various hatted couplings in terms of 
adimensional couplings by means  of the mass parameter $M$, which we take much larger than the  typical energy 
scales that characterize  the IR physics of this model. Therefore :
%%%
\begin{equation}
\widehat a_{s}=\frac{a_{s}}{M^{2(s-1)}} \Bigg |_{s=1,2,3}  \;\;\;\;\;\;\;\; ; \;\;\;\;\;\;\;\;
\widehat g_{n}=\frac{g_{n}}{M^{n-4} }\Bigg |_{n=2,3,4,....}
\label{eq: dimless}
\end{equation}
%%%%
Moreover, in order to set the overall scale of the action, the coefficient of the time derivative of the field in Eq. (\ref{eq: actioneff}) is taken equal to one , 
and, in addition, a relative  rescaling of the space and time  allows us to take $a_1=1$ and, finally, a redefinition of the mass scale $M$ can be 
performed to set also $a_3=1$, while $a_2$ remains an unconstrained coupling.  Thus, we take :
\begin{equation}
\widehat a_{3}=\frac{a_3}{M^{4}}= \frac{1}{M^{4}} \;\;\;\;\;\;\;\; ; \;\;\;\;\;\;\;\; \widehat a_{1}=a_1=1\;\;.
\label{eq: one}
\end{equation}
%%%%

Now we turn to the computation of the quantum corrections. In Section \ref{sec2} we verified that the  divergent 
diagrams generated by this model must contain at most one vertex, which means that we have only  one 
logarithmically divergent integral, namely
%%%%%%%%%%
\begin{equation}
\widehat I_1=\int\!\frac{d^{3} k \; d k_0 }{(2\pi)^{4}}
\; \frac{i}{ \left ( k_{0}^{2} - \widehat a_3  \vec{k}^{\,6}    - \widehat a_2 \, \vec{k}^{\,4} - \widehat a_1  \vec{k}^{\,2} - \widehat g_2  + i \epsilon \right ) }
=
\frac{1}{2}\,
\int\!\frac{d^{3} k }{(2\pi)^{3}}
\; \frac{1}{\sqrt{\widehat a_3  \vec{k}^{\,6}     + \widehat a_2 \, \vec{k}^{\,4} +\widehat a_1  \vec{k}^{\,2} + \widehat g_2  } } 
\label{eq:int0}
\end{equation}
%%%%%%%%%%%
In Eq. (\ref{eq:int0}) we performed the integral in $k_0$ and we are left with the integral over the three-momentum $\vec k$
that will be calculated for $|\vec k |$ between the UV cut-off $\Lambda$  and an IR cut-off  which we choose, for the moment, 
equal to $M$ in order to neglect, in first approximation, other smaller scales such as $\widehat g_2$ :
%%%%%%%%%%
\begin{equation}
\widehat I_1(\Lambda, M)= M^2 \; I_1 (\Lambda, M)= \frac{M^2}{(2\pi)^2} \;\;{\rm ln}\left(  \frac{\Lambda}{M}  \right) +O\left(\frac{M^4}{\Lambda^2}\right)
\label{eq:intlog}
\end{equation}

If we now focus on  the quantum corrections for the quartic coupling $\widehat g_4=g_4$, we notice that the collection of divergent, single vertex diagrams 
includes the sum of the diagram with vertex $\widehat g_6$  and the one loop integral $\widehat I_1$, plus the diagram with vertex $\widehat g_8$  and two loops
corresponding to  the square of the integral $\widehat I_1$,
and so on, until the last diagram with $\widehat g_{\overline n}$  and $\widehat I_1^{(\overline n/2-2)}$. After counting the combinatorial factors 
and by making use of the adimensional couplings, we get the renormalized quartic coupling  $g_{4R}$
%%%%%%%%%
\begin{equation}
g_{4R}= g_{4} + g_{6}\left( \frac{I_1}{2} \right) + \frac{g_{8}}{2} \left( \frac{I_1}{2} \right)^2 + \frac{g_{10}}{3!} \left( \frac{I_1}{2} \right)^3 +....+
\frac{g_{\overline n}}{(\overline n/2 -2)!} \left( \frac{I_1}{2} \right)^{(\overline n/2 -2)}
\label{eq: serie}
\end{equation}
%%%%%%%%%%%%%%
where we assumed $\overline n$ to be an even integer (for odd $\overline n$, one has just to replace $\overline n$ with $(\overline n-1)$ in Eq. (\ref{eq: serie})).

Moreover, the structure of this series of diagram is such that when we compute $g_{6R}$ or $g_{8R}$, we get exactly 
the same series as in Eq. (\ref{eq: serie}), with the only difference that each coupling index is increased of 2 units (for $g_{6R}$) or 4 units (for $g_{8R}$),
while the truncation occurs at $\overline n$ in all cases. With this input we can invert the truncated series (\ref{eq: serie}) to get  $g_{4}$ in terms of
renormalized couplings:
%%%%%
\begin{equation}
g_{4}= g_{4R} - g_{6R}\left( \frac{I_1}{2} \right) + \frac{g_{8R}}{2} \left( \frac{I_1}{2} \right)^2 - \frac{g_{10R}}{3!} \left( \frac{I_1}{2} \right)^3 +....+
\frac{g_{\overline nR}}{(\overline n/2 -2)!} \left( \frac{-I_1}{2} \right)^{(\overline n/2 -2)}
\label{eq: invserie}
\end{equation}
%%%%%%%%
and in this case we find alternating signs. In addition, due to the topology of the divergent diagrams, the same argument can be repeated with equivalent 
conclusions for the couplings with odd index, $g_{3},\,g_{5}, ... ,g_{(\overline n-1)}$.

From the  general relation in (\ref{eq: serie}), we evince the result that 
couplings with even (odd)  index $n$ get divergent corrections only from other 
couplings with even (odd)  index $m$, such that $n< m\leq \overline n$. Then, couplings with $n>\overline n$, 
that do not appear in the bare action (\ref{eq: actioneff}), get only finite corrections from subleading diagrams that contain more than one vertex.
Therefore, the theory is perturbatively renormalizable because all divergences are under control and can be discarded by the conventional insertion
 of  counterterms in Eq. (\ref{eq: actioneff}). 

Depending  on the value of  $\overline n$, and therefore on the number of terms of the sum in Eq. (\ref{eq: serie}),
we  get a larger or smaller correction to $g_{4}$. For instance,  when  $\overline n=6$,  $g_{4}$  gets only a logarithmic correction 
whereas for large $\overline n$, $g_{4}$ gets larger corrections that are proportional to higher  powers of the logarithm, while the 
coupling $g_{\overline n-2}$ gets  linear corrections in the logarithm and $g_{\overline n}$ does not have any divergent correction.

An interesting remark concerns the particular case in which all even couplings are equal ($g_{4} =g_{6} = g_{8} =...=g_{\overline n}$) and
the right hand side of Eqs. (\ref{eq: serie}) becomes the truncated exponential series or, analogously, if $g_{4R} =g_{6R} = g_{8R} =...=g_{\overline nR}$,
Eq. (\ref{eq: invserie})  is the truncated exponential series with alternate signs. 
If we consider the limit $\overline n\to \infty$ under the assumption
of equal couplings, we get from Eq. (\ref{eq: serie})
%%%%%
\begin{equation}
g_{4R}=g_{4} \; {\rm e}^{I_1/2}\;\;,
\label{expo}
\end{equation}
%%%%%%
and the same result is obtained by starting from Eq. (\ref{eq: invserie}). In addition, it is evident that in the limit $\overline n\to \infty$ the relation 
between bare and renormalized coupling in Eq. (\ref{expo}) holds not only for the quartic coupling but for all couplings with even index and,
as a consequence, if the bare couplings are all equal, then also the renormalized couplings are all equal (or viceversa).
Clearly, all these results can be equivalently recovered for the set of couplings with odd index.

Eq. (\ref{expo}) shows that, in order to keep $g_{4R}$ finite when $\Lambda\to \infty$, $g_4$ must vanish in the same limit.
Therefore, a peculiar result  emerges for the  exponential series where the couplings are all equal.
Its extension to the more general case with different couplings $g_n$ can be related 
to the possibility of constraining the given series of $g_n$ with a suitably constructed exponential series which, for instance, is 
larger than the former, term  by term (provided that $g_n>0$ for all $n$). We can conclude that the more general series is  summable, giving 
finite or vanishing  $g_n$ in the limit $\Lambda\to \infty$, and the associated model  is UV safe or free.

Finally, the coefficients of the derivative terms, $a_1\,,a_2\,,a_3,$ get corrections from diagrams that carry some external momentum dependence; 
this necessarily requires the presence of at least two vertices and consequently only UV finite corrections affect these three couplings.
Therefore, the leading corrections come from the one loop diagram with two $g_3$ vertices and two internal lines, corresponding to the one loop 
integral $R_1(p)$ in Eq. (\ref{eq: poles}), and  from the two loop diagram with two  $g_4$ vertices and three internal lines (sunset diagram) 
corresponding to the  integral ($i\epsilon$ in the propagators is omitted)
\begin{equation}
R_2(p)=\int\!\frac{d^{3}k  \, d k_0 }{(2\pi)^{4}}\,
\int\!\frac{d^{3}q  \; d q_0 }{(2\pi)^{4}}\; 
\frac{1}{
 k_{0}^{2} -D(\vec k)^{2} }
\;\;
\frac{1}{q_{0}^{2} -D(\vec q)^{2} }
\;\;
\frac{1}{ (p_0+k_{0}+q_0)^{2} -D(\vec p + \vec k +\vec q)^{2} }
\label{eq: twolodi}
\end{equation}
where, similarly to Eq. (\ref{eq:definitions}), we define 
$D(\vec k)=\sqrt{\widehat a_3 \, \vec{k}^{\,6}     + \widehat a_2 \, \vec{k}^{\,4} + \widehat a_1 \, \vec{k}^{\,2} + \widehat g_2}$.
Then, explicitly,  the leading corrections  to $a_1\,,a_2\,,a_3,$ are obtained from the coefficients of the expansion of these diagrams in 
powers of the external momentum $\vec p$
($j=1,2,3$) 
\begin{equation}
\label{acorr}
\delta a_{j} = \frac{1}{(2j)!}\;\frac{\partial^{2j}}{(\partial |\vec p |)^{2j}} \; \Bigg [ \frac{i \;{\widehat g_3}\,^2 }{2} R_1 -  \frac{ {\widehat g_4}\,^2 }{6} R_2 \Bigg]\; \Bigg |_{p=0}
\end{equation}

The approximation in Eq. (\ref{acorr}) represents the leading correction to the parameters $a_j$, 
both when $k\equiv |\vec k |>>M$ (and  the propagators are dominated by the term  $(a_3\, k^6/M^4)$), and when $ k<<M$ (and the dominant contribution  
to the propagator is now proportional to $a_1\, k^2$,  while  terms proportional to $a_3$ and  $a_2$ are suppressed).
In fact,  further corrections to $a_j$ involve more vertices $\widehat g_n$ with  $n>4$,  and a higher number of loops.
They are UV finite,  according to the analysis of Sec. \ref{sec2}, and smaller in size than the leading corrections because of the numerical factor 
carried by each loop. Therefore, we do not include them in Eq. (\ref{acorr}) where, in addition, we keep $\widehat g_3$ and   $\widehat g_4$ constant,
omitting higher order corrections related to the couplings. This implies that $\delta a_j$ in (\ref{acorr})
are independent of the specific value of $\overline n$ in Eq. (\ref{eq: poten}).

An inspection of Eq. (\ref{acorr}) in the UV region (i.e. when the momentum in the integrals $R_1$, $R_2$ is integrated
from $M$ to $\mu>>M$)
shows the same power law behavior for the $O( \widehat g_3\,^2)$ and $O( \widehat g_4\,^2)$ corrections:
%%%%%%%
\begin{equation}
\delta a_{j}\propto \left(\frac{M}{\mu}\right )^{6+2j}
 \;\;.
\label{finitecorr}
\end{equation}  
%%%%%%%
Then, in this region
$a_1\,,a_2\,,a_3$, get only negligible corrections and, for practical purposes, they maintain the value  assigned in Eq. (\ref{eq: one}) 
(in particular $a_1=a_3=1$). 

In the  IR region, i.e. when  momentum is integrated from zero to  $\mu << \sqrt {\widehat g_2} <M$,
(provided $\widehat g_2>0$), all 
$D$, defined after Eq.(\ref{eq: twolodi}),  included in the propagators of $R_1$, $R_2$ in  (\ref{acorr}),
 are essentially proportional to   $\sqrt {\widehat g_2}$,  so that
the contribution to $\delta a_j$ is proportional to powers of  $(\mu^2/\widehat g_2)$, and therefore negligible.

The picture  of the  intermediate momentum region (with the integration variable ranging  from $\mu \simeq \sqrt {\widehat g_2}$
 to $M$),  is less evident, since more than two scales are involved in the computation of   $R_1$ and $R_2$,
and a numerical analysis of Eq.(\ref{acorr}) is in order. This analysis, discussed in  Sec. \ref{sec4}, 
shows a negative correction $\delta a_3<0$ from the intermediate momentum region, with
$|\delta a_3|$  proportional to the square couplings $\widehat g_3^2$ and $\widehat g_4^2$,  as deducible from Eq. (\ref{acorr}).
It follows that there are particular conditions (a too small -or negative- $\widehat g_2$ and large couplings 
$\widehat g_3$,  $\widehat g_4$)  in which $\delta a_3<-1$ and, after including these corrections in the factors $D$ in 
Eq.(\ref{acorr}), we end up with the square root of negative numbers. This singular behavior, appearing at IR scales 
close to $\sqrt {\widehat g_2}$, can be  avoided by taking  smaller $\widehat g_3$ and $\widehat g_4$, and/or larger $\widehat g_2$.

In addition, simple dimensional analysis of Eq.(\ref{acorr})  shows that $|\delta a_2|$ and $|\delta a_1|$ are smaller than $|\delta a_3|$,
respectively  by the factors $\mu^2/M^2$ and $\mu^4/M^4$.
%%%%%%%%%%%%%%%%%%%%%%%%%%%%%%%%%%%%%%%%%%%%%%%%%%%%%%%%%%%%%%%%%%%%
\section{Flow of the couplings}
\label{sec4}
%%%%%%%%%%%%%%%%%%%%%%%%%%%%%%%%%%%%%%%%%%%%%%%%%%%%%%%%%%%%%%%%%%%%%%

By exploiting the dependence of the integrals on the renormalization scale, we easily transform  Eq. (\ref{eq: serie})  and the analogous relations 
for all the other couplings, into a set of differential flow equations for the scale dependent couplings $g_{n}(\mu)$, and study the evolution of the parameters with 
$\mu$ going toward the IR region at fixed boundaries at a large UV scale or, equivalently, explore the UV behavior  by fixing the  boundaries at a lower scale. 
To this purpose, we express all dimensionful quantities in terms of $M$ and display the flow of the running $g_{n}(\mu)$ as a function of 
the logarithm of the scale $\mu$:  $t={\rm ln }(M/\mu)$. 

\begin{figure}
\begin{centering}
\includegraphics[width=9.5 cm,height=7cm]{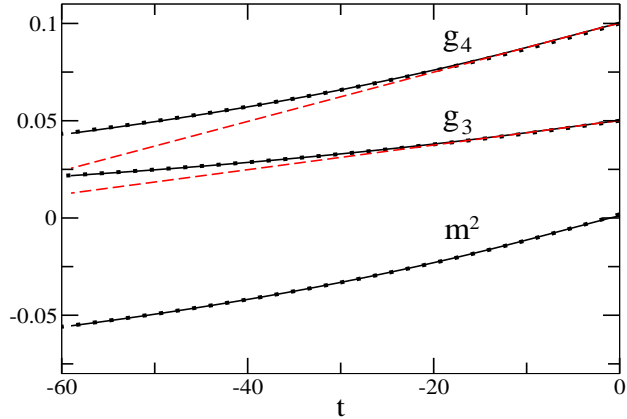}
\par\end{centering}
\caption{\label{figure1} UV flow of the couplings $g_{3}, g_{4}$, with boundaries fixed at $t=0$, for two  different values
of $\overline n$, namely $\overline n=22$, black solid lines, and  $\overline n=6$, red (online) dashed lines.
At $t=0$, for all $g_{n}$ with $n$ even and $4\leq n \leq\overline n$, the  boundary value  $0.1$ is used, while $0.05$ is used for odd $n$.
The flow of $m^2=g_{2}$ is also displayed for $\overline n=22$ with boundary value $10^{-3}$ for $m^2$ and 
the same boundaries as before for the other couplings. Dotted black lines correspond to the exponential approximation, 
as  in Eq. (\ref{expo2}), to the various curves.}
\end{figure}

We start by investigating the UV region above $M$,  where the effect of the modified propagator should be evident. We focus on some representative couplings,
namely  $g_{3}$ and $g_{4}$  plus the square mass term $m^2=g_2$, as in this region  $a_1\,,a_2\,,a_3$, that contribute to the propagator, 
remain practically constant.
In Fig.\ref{figure1} the UV behavior of the couplings $g_{3}$ and $g_{4}$ is studied by fixing the boundary conditions with all even couplings equal to $0.1$ and
all odd couplings equal to $0.05$  at $t=0$ (i.e. at $\mu=M$), for two different values of $\overline n$ in Eq. (\ref{eq: poten}), 
namely $\overline n=22$ (black solid curves) and $\overline n=6$ (red dashed curves). 
Note that, according to the definitions given in Eq. (\ref{eq: dimless}), all values of the couplings are expressed in units of $M$.

The linear dependence of the red dashed curves is clearly evident. It is due to the choice $\overline n=6$ which implies the contribution of 
a single  one loop diagram to the flow of $g_3$ and $g_4$. The different slope of the two dashed curves 
is instead due to the different values of the vertices associated to the one loop diagram respectively for  $g_3$ and $g_4$.
Conversely, the solid black lines are obtained for a very large value of $\overline n$, and we chose the same initial value of all even couplings and another
initial value for all the odd couplings, with the aim of pointing out the exponential nature of the truncated series in Eq. (\ref{eq: serie}). This trend
is confirmed by the dotted black lines that reproduce  the plot of the resummed exponential series shown in  Eq. (\ref{expo}) which, 
in the present notation, becomes (for $j=3,4$)
%%%%%
\begin{equation}
g_{j}(t)=g_j(t=0) \; {\rm e}^{\,\frac{t}{8\pi^2}} \;\;.
\label{expo2}
\end{equation}
%%%%%%
As already stated, at fixed $g_j(t=0)$, the running couplings $g_j(t)$ vanish  in the limit  $t\to -\infty$.

Both dashed and solid lines are negatively divergent,  the former linearly, as already noticed above, 
while the latter does actually diverge as a power of $t$ due to the last term in the right hand side of Eq. (\ref{eq: serie}) :
$\sim (- g_{\overline n})\; (-t)^{(\overline n/2 -2)}$  (according to our choice on the values of the couplings, 
$g_{\overline n}$ is the last non-vanishing coupling and  does not get any correction).  

Nevertheless, the divergent trend of the latter becomes  evident only at extremely large negative values of $t$, not contained in Fig.\ref{figure1},   
as an effect of  the sum over the large number ${\overline n}$ of terms.  On the contrary, Fig.\ref{figure1} manifestly indicates that  
the exponentially vanishing  expression in Eq. (\ref{expo2}) provides an excellent approximation to  the actual flow of our couplings 
up to  $t=-60$  (i.e.  $\mu/M\sim 10^{26}$) at least for the choice of the  initial values of the couplings considered above.

If we select ${\overline n}=24$ or ${\overline n}=8$ in Fig.\ref{figure1}, instead of ${\overline n}=22$ or ${\overline n}=6$,
then  the dotted  and dashed curves  are modified  because, according to Eq. (\ref{eq: invserie}), the couplings now 
tend to $+\infty$ as a power of $|t|$, when $t\to -\infty$, and a change of slope  at some $t<0$ with the generation of a  minimum is observed.
However, as discussed above, this change  occurs at large  $|t|$, the flow of $g_{4}$ and $g_3$
 being dominated respectively  by $g_{6}$ and $g_5$ at small $|t|$. In particular, for  ${\overline n}=24$ and
the initial values used in  Fig.\ref{figure1},  the trend  of Eq. (\ref{expo2}) is preserved 
for many orders of magnitude, up to $t=-60$, before  the effects of $g_{24}$ become apparent.

In Fig.\ref{figure1}  the square mass $m^2=g_2$ is also plotted (solid line) for the case with  $\overline n=22$ and the boundary values of other 
couplings as declared above. The corresponding exponential curve is also reported (dotted line).
Actually, the square mass in the UV region where the relevant scales are much larger than the mass itself,  has the same of trend of the other
couplings. Therefore, it is not surprising that its flow is totally similar to those already examined. The only difference  is $m^2 (t=0)=10^{-3}$,
much smaller than the value of the other two couplings. In fact, we expect the renormalized mass of such a theory to be well below the 
Lorentz violating reference scale $M$,  i.e.  we expect  $m^2<<1 $ at $t=0$. 

\begin{figure}
\begin{centering}
\includegraphics[width=9.5 cm,height=7cm]{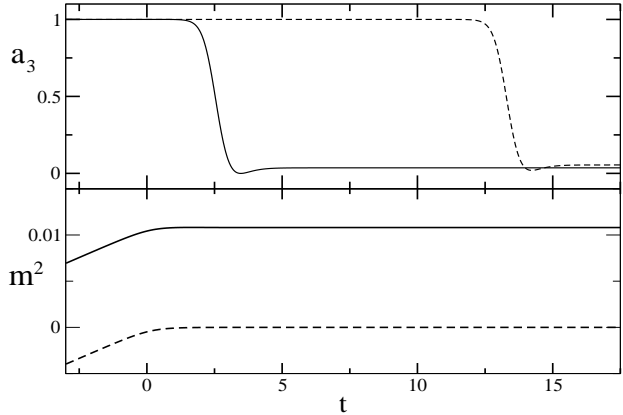}
\par\end{centering}
\caption{\label{figure2} Flow of the parameter $a_3$ (upper panel) and of the square mass $m^2=g_2$ in the region around $t\sim 0$ and
in the IR region with $t>0$. Boundary values are taken at $t=-3$, with  $m^2=0.007$ and $g_3=g_4=0.1$ in one case (solid lines) 
and  $m^2\simeq -0.004$, $g_3= 10^{-15}$, $g_4=0.1$ in the other case (dashed lines).
As clarified in the text, diagrams containing
vertices $g_n$, with $n>4$, are neglected in this case and all plots do not depend on $\overline n$ in (\ref{eq: poten}). 
}
\end{figure}

Let us now consider the momentum region around the scale $M$, corresponding to the transition from UV 
to  IR regime in which, due to the suppression of terms proportional to powers of $k/M$ in the propagator, the onset of the well known 
standard scaling occurs. 
While no sensible change in the three coefficients $a_1$, $a_2$, $a_3,$ is detectable in the UV region
according  to the explanation in Sec. \ref{sec3}, at momenta below $M$ we expect a 
change in $a_3$  due to the finite corrections induced by the one and two loop diagrams, as shown in Eq.(\ref{acorr}).
However, for the sake of simplicity, in the following computation of the evolution of $a_3$, only the one loop diagram proportional to $g_3^2$
is included, while we neglect the two loop diagram proportional to $g_4^2$, subdominant with respect 
to the one loop diagram, because of the higher loop order. 

The location of the change in $a_3$ is strictly related to the particular value of the square mass $m^2$. Therefore we report in the same figure
both running parameters, $a_3$ and $m^2$, namely the former in the upper panel and the latter in the lower panel of Fig.\ref{figure2},
for two particular choices of the mass (solid and dashed lines in both panels). 
The boundary conditions are fixed at $t=-3$, and in the first case, that corresponds to the solid lines in Fig.\ref{figure2},
we take $g_3=g_4=0.1$ and square mass $m^2=0.007$. 
The leading contribution to the flow of $m^2$, both above and below the scale $M$, comes from the one loop diagram proportional to $g_4$,
whose effect  is to increase the mass for growing $t$, although with different rates for $t<0$ or $t>0$
(in fact also the $O(g_3^2)$ one loop diagram is included but its contribution to the mass correction is practically negligible). 
Then, for large enough $t$, when the running scale  becomes lower than the mass $m$, the flow of all couplings stops
 and this is clearly visible in all plots of Fig.\ref{figure2}.
 
According to all phenomenological indications, we expect the scale $M$ to be various orders of magnitude larger than 
the IR mass of our model and this, despite the much softer behavior of the parameters in the UV region,  inevitably requires an 
accurate adjustment of the UV boundary value of $m^2$.  So, unlike the first example in Fig.\ref{figure2},  where at large $t$ we have $m^2\simeq 10^{-2}$,
in the second example corresponding to dashed curves, 
we tune the boundary value of $m^2(t=-3)$ at a negative value, and $g_3=10^{-15}$,
to get in the IR region $m^2\simeq 10^{-12}$. 
(Actually, the latter IR value of $m^2$ is chosen as  a compromise between dealing with a very  small 
but still numerically manageable number, and having a not too large physical mass,
whose scale is set by $M$, which can be as large as, or  larger than the Plank mass).

Figs.\ref{figure1}, \ref{figure2} show that our model produces a negatively growing  $m^2$ when $t={\rm ln}\left({M}/{\Lambda} \right) \to -\infty$.
Something similar occurs in the Lorentz invariant renormalizable $d=4$ scalar theory  but, while the 
latter provides a  quadratically divergent $m^2\propto -\Lambda^2$, in our model  
the quadratic trend occurs only below  $M$ (hardly visible in Fig.\ref{figure2}  because of the units  chosen) while 
above $M$, it turns into the logarithmic trend  shown in Fig \ref{figure1}.

In the upper panel of Fig.\ref{figure2}, the flow of $a_3$ is shown and a  comparison of  the solid and dashed lines indicates
that the integration of the intermediate momentum region in Eq. (\ref{acorr}) produces a rapid decrease of $a_3$, in correspondence of the 
the scale $t={\rm ln}\left( {M}/{m} \right)$, where $m=m(t)$  is the corresponding value of the running mass, readable from the lower panel, 
i.e.  $m\simeq \sqrt{0.01}$ for the solid line, and $m\simeq \sqrt{10^{-12}}$ for the dashed line.
Above and below these crossover scales, both solid and dashed curves are essentially flat, indicating that  $\delta a_3\simeq 0$ in those regions, 
as suggested by the general analysis of Eq. (\ref{acorr}) in Sec. \ref{sec3}.

However, as explained in Sec. \ref{sec3}, the drop of $a_3$ in Fig.\ref{figure2} is proportional to the square coupling 
($g_3^2$ in the present example) and if the running  mass is too small, this can generate a negative argument of the square root
in the factors $D$ appearing in $R_1$ and $R_2$ in Eq. (\ref{acorr}). To avoid this singular behavior, we choose a sufficiently large 
mass in the case of the solid line, while in the case of the dashed line where, as discussed above, we tuned the mass to the value
$m\simeq 10^{-6}$, we must select the small value $g_3\simeq 10^{-15}$.

As anticipated in the analysis of Eq.(\ref{acorr}) in Sec. \ref{sec3},  the corrections $\delta a_2$ and $\delta a_1$ of the other two derivative couplings 
are suppressed  with respect to $\delta a_3$. In fact, for the two examples discussed in Fig.\ref{figure2}, the numerical determination of the flow yields 
deviations  from the boundary values $a_1=1, a_2=a_2^B$ (as discussed before Eq. (\ref{eq: one}), $a_2$ is left unconstrained)
that are small  (especially $\delta a_1$) and of no practical relevance in this investigation and, therefore, negligible.
Moreover, all numerical  findings show small dependence on 
the specific value of $a_2^B\simeq O(1)$,   so that, for the sake of simplicity, we take  $a_2^B=1$.

\begin{figure}
\begin{centering}
\includegraphics[width=9.5 cm,height=7cm]{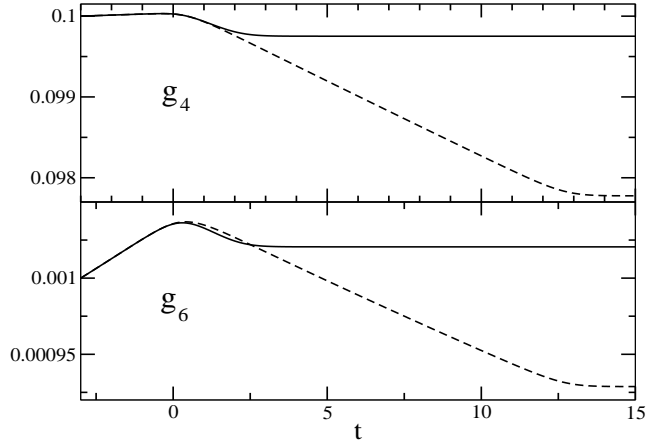}
\par\end{centering}
\caption{\label{figure3} 
Flow of $g_4$ (upper panel) and $g_6$ (lower panel) with the same boundary values (and the same coding) adopted in the two cases 
displayed in Fig.\ref{figure2}, with further  boundary values: $g_6=g_8=0.001$ at $t=-3$. }
\end{figure}

Finally, in Fig.\ref{figure3} the crossover of the couplings $g_4$ (upper panel)  and  $g_6$  (lower panel) from one regime to the 
other is shown with the boundaries taken as in the two cases already discussed  in Fig.\ref{figure2}. So, solid lines correspond to 
large positive square mass at $t=-3$, while dashed lines to $m^2(t=-3)<0$ and $g_3\simeq 10^{-15}$.
In both cases we fixed $g_4=0.1$ and $g_6=g_8=0.001$ at $t=-3$. 
In addition,  we do not include diagrams containing vertices $g_n$   with $n>8$ which, as discussed below, in this momentum region
are finite and of higher order in the pertubative expansion, and therefore numerically negligible. 
This in turn implies that, similarly to $\delta a_j$ in (\ref{acorr}), the flow of $g_4$ and $g_6$
in this region is independent of the specific value of $\overline n$ in Eq. (\ref{eq: poten}).

Both $g_4$ and $g_6$ show similar trends: increasing   for  $t<0$ 
(due to the different scale in the two panels, the slope of $g_4$  is less evident although equal to that of $g_6$)
and decreasing for $t >0$ and finally flat at larger $t$, when the momentum scale  becomes smaller than the mass $m$. 
The change of slope around $t=0$ is a clear indication of the different influence of various diagrams in the two regions. 

In fact,  when  $t<0$, as repeatedly discussed,  the relevant contribution comes from  the single vertex, one loop diagram, which is $O(g_6)$ in the 
flow of the coupling $g_4$ and  $O(g_8)$ in the flow of $g_6$ (note that we choose $g_6=g_8$ in Fig.\ref{figure3}).

When $t>0$,  the term $\widehat a_1 k^2=a_1 k^2$ is dominant with respect to 
$\widehat a_2 k^4=a_2 k^4/M^2$ and to $\widehat a_3 k^6=a_3 k^6/M^4$ in the propagators of the theory, and it is easy to realize
by simple dimensional analysis that  the one loop diagrams relevant when $t<0$ and mentioned above, are now suppressed by the factor $g_6 \,e^{-2t}$ in the
flow of $g_4$, and by the factor $g_8 \,e^{-2t}$ in the flow of $g_6$. The same dimensional analysis also shows that the $p$-loop diagrams resummed  
(because of the same order)  in Eq.(\ref{eq: serie}) in the flow of $g_4$  when $t<0$, are suppressed by the factor $g_{4+2p} \,e^{-2pt}$ in the same flow
when $t>0$. Therefore, they are less and less important when $n$ grows, and a resummation like the one in Eq.(\ref{eq: serie}) is meaningless when $t>0$.

On the contrary, when $t>0$, according to standard perturbation theory,
the dominant one loop diagrams have two vertices, as signaled by the the change of slope of the curves.
Namely, they are the $O(g^2_4) $ diagram for the flow of $g_4$ and the $O(g_4\,g_6)$ diagram for  $g_6$, 
which are those utilized in the computation 
of the flows  in  Fig.\ref{figure3}.

%%%%%%%%%%%%%%%%%%%%%%%%%%%%%%%%%%%%%%%%%%%%%%%%%%%%%%%%%%%%%%%%%%%%
\section{Comments and conclusions}
\label{sec5}
%%%%%%%%%%%%%%%%%%%%%%%%%%%%%%%%%%%%%%%%%%%%%%%%%%%%%%%%%%%%%%%%%%%%%
The first conclusion that emerges from  the above analysis is that the specific higher derivative scalar theory 
considered in this paper does not produce Lorentz violating effects in the low energy sector, 
large enough to be in contradiction with  experimental observations. In fact, the action in Eq. (\ref{eq: actioneff}) 
generates quantum corrections that are strongly suppressed at high momentum because of the modified propagator,
and its anisotropy parameter $z$ is chosen to be $z=3$, the highest possible value at which the  Lifshitz scaling 
protects the UV asymptotic behavior of the theory. 

However, this is not sufficient to avoid dangerous corrections 
to  $a_1$ that are responsible for the too large Lorentz violating effects, already observed in \cite{Iengo} 
for theories with $z=2$. To avoid these effects, it is necessary to remove from the bare action all momentum dependent vertices, thus reducing precisely 
to the action in Eq. (\ref{eq: actioneff})  where the vertices, which  are  all included in the potential $V$, do not contain any derivative of the field.
In other words, we have to restrict our analysis in the parameter space around the fixed point solution (\ref{eq: lpaction}), to the manifold 
in which all couplings associated to momentum dependent vertices are turned off. Clearly in the present form, this selection rule is not
a consequence of some physical symmetry, 
but nevertheless, as explained in Sec. \ref{sec2},  
it is self-consistent, in the sense that no UV divergent terms are generated, of the same kind of  those not included in the bare action, 
and, moreover,  it is sufficient to avoid those experimentally unacceptable effects.

In fact, the  bare action (\ref{eq: actioneff}) can be normalized by taking
 $a_1=a_2=a_3=1$, and then, the corrections found for  $a_3$ are of the kind shown in the upper panel of Fig.\ref{figure2},
 with the drop from $a_3=1$ to $a_3 \simeq 0$ occurring  at a scale close to the mass $m$,
while for $a_1$ and $a_2$ no significant correction is observed.
Therefore, in the dispersion relation derived from Eq. (\ref{eq: actioneff}) (which has the same form introduced in \cite{Coleman:1998ti})
%%%%%%%%%
\begin{equation}
E^2={\vec k}^{\,2} \, \left [ a_1+ a_2 \left (\frac{k}{M} \right)^2 +    a_3 \left (\frac{k}{M} \right)^4      \right] \,+\,m^2 \; ,
\label{dr}
\end{equation}
%%%%%%%% 
the suppression of the Lorentz violating terms  in the relevant momentum region, $M>> k > m$, is essentially due to the  factors  $(k/M)^2$, 
and  $(k/M)^4$,  as no logarithmic correction affects $a_1$ and, in practice, in that region we still have  $a_1=a_2=a_3=1$. 
If we accept  the loose assumption  that the same dispersion relation can be extended to  more realistic theories and in particular to photons  (but with $m=0$), 
then, the experimental observations allow us to push the scale $M$ around or above the lower limit $M> 10^{17}$ GeV \cite{Ellis:2018lca}; however,
conservatively, we have  to retain this constraint just  as a broad  indication rather than a rigorous lower limit on $M$.

Concerning the renormalization  of the theory with fixed $\overline n$ in Eq. (\ref{eq: poten}), the divergent diagrams produced are all of the 
same kind, with logarithmic (and no quadratic) divergences. Therefore, the renormalization  procedure is easy to handle and all divergences can be cured 
by means of the  introduction of counterterms.

On the other hand,  we can regard the  action (\ref{eq: actioneff})   as an effective theory with range of validity below some large UV  momentum 
scale $\Lambda>>M$,  and such that  it reproduces the well known properties of a standard scalar theory in the IR momentum region  below $M$.
Then, the analysis of the RG flow indicates that, due to the presence of the Lifshitz point,  this theory shows very small changes of the couplings in an
extremely  large range of the running scale $\mu$ (see Fig.\ref{figure1}). Not only
all couplings are not divergent at some Landau pole below  the scale $\Lambda$,  but also they are, in practice, almost constant quantities.

However,  since $g_2$, which is the square mass of the theory,  shows, as the other couplings, only  a small logarithmic change 
in the UV region well above $M$, but gets strong ( O($\mu^2)$ ) corrections below $M$ (these are the large corrections
which give rise to the naturalness problem),  we must conclude that the smoothening of the flow above $M$ does not influence the large 
changes in $g_2$ occurring at lower scales, so that the complication of fine tuning the UV value of the scalar mass is still present.

Finally, the case of very large $\overline n$ with almost equal values of the couplings needs to be emphasised.
In fact, in Section \ref{sec3} we noticed that the limiting theory where all couplings are equal,  turns out to be asymptotically free 
with couplings that vanish when $\mu \to \infty$.
(Incidentally,  similarly to these findings, in \cite{Barv1,Barv2,Barv3,Barv4}
it is shown that the projectable  Horava gravity model  is renormalizable and  asymptotically free both in 2+1 and 3+1 dimensions).
It must be noticed that this limiting case  shows a  potential which is essentially  equal to the one 
of the Liouville theory   (and consequently the properties of the  higher derivative four dimensional  version of the Liouville theory certainly 
deserve further investigation), but at the same time, it can be used to bound more general theories. 

In fact, if we treat  Eq. (\ref{eq: actioneff})  as an effective theory  limited by  an UV cutoff $\Lambda$, then
the boundary condition at $\Lambda$, consisting of very similar (or at least of  the same order of magnitude) 
small values for  a huge number (or even an infinite) of  adimensional couplings,  could be accepted  as a natural assumption. 
Then, a comparison with the Liouville potential should help in establishing constraints on the large momentum (but smaller than $\Lambda$) 
trend of the couplings of the effective theory under investigation and, consequently, in testing whether the latter 
 behaves effectively  as an asymptotically  free (or safe) theory.

\vspace{1cm}

{\it Acknowledgements }: This work has been carried out within the INFN project FLAG.

\end{document}